\documentclass[prl,letterpaper,twocolumn,showpacs,superscriptaddress]{revtex4}
\usepackage{graphicx,psfrag,amsmath,amssymb,amsfonts,bbm,latexsym,color,dcolumn}

\begin{document}

\title{ Dynamical suppression of 1/$f$ noise processes in qubit systems }

\author{Lara Faoro}
\email{faoro@isiosf.isi.it }
\affiliation{Institute for Scientific Interchange Foundation, Viale Settimio 
Severo 65, 10133 Torino, Italy}
\author{Lorenza Viola}
\email{lviola@lanl.gov} 
\affiliation{Los Alamos National Laboratory, Mail Stop B256, 
Los Alamos, New Mexico 87545, USA}

\date{September 18, 2003}

\begin{abstract}
We investigate the capability of dynamical decoupling techniques to reduce
decoherence from a realistic environment generating 1/$f$ noise.  The
predominance of low frequency modes in the noise profile allows for 
decoherence scenarios where relatively slow control rates suffice
for a drastic improvement.  However, the actual figure of merit is 
very sensitive to the details of the dynamics, with decoupling
performance which may deteriorate for non-Gaussian noise and/or 
high frequency working points.  Our results are promising for robust 
solid-state qubits and beyond. 
\end{abstract}


\maketitle

Noise processes characterized by a 1/$f$ power spectrum 
are ubiquitously encountered in Nature.  While a unified theory of the 
underlying mechanisms remains elusive, 1/$f$ noise plays a prominent 
role in dynamical phenomena as diverse as transport in electronic 
devices \cite{Weissman1988}, light emission from astrophysical 
sources \cite{Press1978}, statistics of DNA sequences \cite{Voss1992}, 
and stock market prices \cite{Lo1991}.  In recent years, the continuous 
advances witnessed by device nanotechnologies, along with the challenge 
to implement quantum information processing (QIP) in solid-state systems 
\cite{SpecialIssue}, have sharpened the demand for a detailed 
understanding of 1/$f$ noise effects and for viable compensation 
schemes at the quantum level.  In particular, 1/$f$ noise due to 
fluctuating background charges (BCs) severely hampers the performance 
of single-electron tunneling devices \cite{Covington2000} and Josephson 
qubits in the charge regime \cite{Nakamura1999}.

Prompted by the experimental demonstration of a coherent charge echo in a 
Cooper-pair box \cite{Nakamura2002}, efforts are underway to explore the
possibility of 1/$f$ noise reduction via active control techniques.  Recent 
theoretical analyses \cite{Martinis,Shiokawa2002} largely rely on deriving 
1/$f$ noise from the influence of an harmonic oscillator bath e.g., 
through spin-boson models with sub-Ohmic damping \cite{Shnirman2002}.  
While this accurately represents environments consisting of a {\em continuum} 
of weakly coupled modes, the inherent Gaussian distribution of the fluctuations 
fails at reproducing the distinctive properties of 1/$f$ noise due to 
realistic {\em discrete} environments \cite{Paladino2002}.  Compensation of 
non-Gaussian random telegraph noise (RTN) from a {\em single} bistable source 
is considered in \cite{Gutmann2003}.  However, an appropriate distribution 
of characteristic time scales is required to obtain genuine 1/$f$ noise 
effects \cite{Dutta1981}, thus qualitatively changing the nature of the 
control problem.

In this Letter, we present a comprehensive study of the effectiveness 
of decoupling methods \cite{HaeberlenBook,Viola1} at suppressing 
1/$f$ noise in a single qubit.  
Our approach fully captures both Gaussian and non-Gaussian 
effects for realistic noise spectra and arbitrary operating points of 
the qubit.  We find that the control performance depends critically on 
the {\em frequency location of the dominating 1/f noise sources}, the 
quality of the attainable suppression becoming comparatively higher as 
the latter shifts toward lower frequencies.  For purely Gaussian 1/$f$ 
dephasing, this implies the potential of significant coherence recovery 
(up to 75\%) by using control rates which can be {\em orders of magnitude 
slower} than expected from the fastest characteristic time scale present 
in the system.

{\it Noise model.$-$} A simple way for generating a 1/$f$ spectrum is 
via an ensemble of $M$ independent, randomly activated 
bistable processes.  Let $\xi_k (t)$ be an asymmetric RTN signal 
switching between values $\pm v_k /2$ with rates $\gamma_k^{(\pm)}$, 
$\gamma_k =\gamma_k^{(+)}+ \gamma_k^{(-)}$.  If a distribution $P(\gamma) 
\propto 1/\gamma$ is assumed for the switching rates $\gamma_k \in 
[\gamma_{min}, \gamma_{max}]$ \cite{Dutta1981,Weissman1988}, 
the total fluctuation $\Xi (t) = \sum_k \xi_k (t)$ exhibits a 1/$f$ power 
spectrum of the form $S(|\omega|) = A/|\omega|$, $A >0$, in a frequency 
range defined by effective cutoffs $\gamma^e_{min} \gg \gamma_{min}$, 
$\gamma^e_{max} \ll \gamma_{max}$.  We focus on 
distributions of strengths $v_k$ sufficiently peaked around their
mean value $\langle v \rangle \geq 0$, in which case $A$ is proportional 
to the number $n_d$ of fluctuators per noise decade, weighted by 
$\langle v \rangle^2$.  

The above phenomenological model applies to a variety of 1/$f$ noise 
processes e.g.,  due to delocalized charge traps or hopping defects 
in solid-state devices \cite{Dutta1981}.  In a fully quantum description, 
the RTN ensemble is replaced by an environment $E$ consisting of $M$ 
two-state BCs \cite{Bauernschmitt1993}, each coupled with strength 
$v_k$ to the system and with strength $T_k$ to an electronic band 
inducing relaxation with rate $\gamma_k$.  Thus, $H_E = \sum_k H_k$, with 
\begin{equation}
H_k = \epsilon_k b_k^\dagger b_k + T_k \sum_l [c^\dagger_{kl} b_k +
\text{h.c.}] +\sum_l \varepsilon_{kl} c^\dagger_{kl} c_{kl} \:,
\label{quantum}
\end{equation}
where $\epsilon_k, \varepsilon_{kl}$ are energy parameters and
$b_k (b^\dagger_k)$, $ c_{kl} (c_{kl}^\dagger)$ canonical fermion 
operators, respectively.  The semiclassical approximation is accurate in 
a regime where each BC is strongly coupled to the corresponding band, 
$T_{k} \gg v_k$, implying the possibility to neglect quantum back-action 
effects from the system.  Let $g_k=v_k/\gamma_k$ for each noise source.  
The resulting decoherence contribution is qualitatively 
different depending on whether $g_k \ll 1$ (Gaussian source) or not 
\cite{Paladino2002}.  While purely Gaussian noise may be formally 
reproduced by an appropriate bosonic bath \cite{Shnirman2002}, 
non-Gaussian 1/$f$ behavior may exhibit non-equilibrium saturation 
features and pronounced non-Lorentzian lineshapes 
\cite{Paladino2002}.

{\it Decoupling setting.$-$} Our target system $S$ is a single qubit,
described by a Hamiltonian $H_S= -[{\Omega} \sigma_z  + {\Delta} \sigma_x] 
/2$, $\sigma_z$-eigenstates $\{ |0\rangle, |1\rangle\}$ defining the 
computational basis, and $\Delta E= \sqrt{ \Omega^2 + \Delta^2}$ giving 
 the energy scale of the free dynamics ($\hbar=1$).  For the charge 
qubit realized in \cite{Nakamura1999,Nakamura2002}, $\Omega$ is the 
separation between the two lowest charge states differing by a 
single Cooper pair, while $\Delta$ is the Josephson energy of 
the junction.  Coupling with the BCs is introduced via an interaction 
$H_{SE}=$$\sum_k v_k b^\dagger_k b_k \otimes \sigma_z $, corresponding 
semiclassically to the effective Hamiltonian $H_{RTN}(t)=\Xi(t)\, \sigma_z$.  
Both relaxation and dephasing dynamics occurs, depending on the longitudinal 
or transverse nature of the fluctuations relative to the physical 
basis defined by the energy eigenstates.  When $\Delta=0$, the physical
and computational bases are aligned, and decoherence is purely adiabatic.  
In the opposite case where $\Omega=0$, purely non-adiabatic dissipation 
and dephasing take place, both mechanisms influencing decoherence in the 
$z$ basis as the latter is now 90$^\circ$ displaced.  Borrowing from the 
Cooper-pair-box terminology, we shall refer to these limiting situations 
as pure dephasing and charge degeneracy regimes, respectively 
\cite{Paladino2002}. 

Dynamical decoupling strategies coherently average out the effects 
of unwanted interactions over a sufficiently long time scale by means 
of a tailored control field \cite{Viola1}.  In the simplest formulation, 
decoupling is achieved by subjecting the system to cyclic sequences 
of instantaneous (bang-bang) control pulses.  We consider two elementary 
decoupling protocols specified by the following 
control cycles: ${\cal P}_A=\{\Delta t , \pi_x , \Delta t , \pi_{-x}\}$
(asymmetric) and ${\cal P}_S=\{\Delta t/2 , \pi_x , \Delta t , 
\pi_{-x}, \Delta t/2 \}$ (symmetric), where $\Delta t$ and 
$\pi_{\pm x}$ denote a free evolution interval of duration 
$\Delta t$ and a controlled rotation of 90$^\circ$ about the $\pm x$ 
axis, respectively.  While the two implementations are equivalent
in the ideal limit of arbitrarily fast control $\Delta t \rightarrow 0$, 
superior averaging is expected from the time-symmetric protocol in the 
realistic case of finite $\Delta t$ thanks to the cancellation of 
higher-order corrections \cite{HaeberlenBook,Viola1,Gheorghiu}. 
 
{\it Decoupled dynamics.$-$} Complete information about the 
dissipative qubit dynamics is contained in the expectation values of 
the qubit observables $\sigma_\ell$, $\ell=x,y,z$, after evolution 
in the presence of the 1/$f$ disturbance with and without the decoupling 
field.  For a generic $H_S$, the latter have been calculated numerically
as the ensemble average over the stochastic qubit state evolved under
$H_{RTN}(t)$ \cite{FaoroNext}, 
$\langle \sigma_\ell (t) \rangle ={\cal E} \{ \langle \psi(t) | 
\sigma_\ell | \psi(t) \rangle \}$, starting from known pure-state 
initial conditions $|\psi (0)\rangle= c_0 |0\rangle + c_1 |1\rangle$, 
$|c_0|^2+|c_1|^2=1$.
We consider here two performance indicators: $\langle \sigma_+ (t) 
\rangle = \langle \sigma_x (t) + i \sigma_y (t) \rangle/2$, 
starting from $c_0=c_1=1/\sqrt{2}$, which describes coherence in the 
$z$ basis;  and $\langle \sigma_z (t) \rangle$, starting from $c_0=1, 
c_1=0$, which is a signature for coherent oscillations in the $x$ basis.  
At $\Delta =0$, an analytic benchmark has been obtained by both performing 
the semiclassical approximation in the Heisenberg equations of motion 
derived from Eq. (\ref{quantum}) and by exactly solving the RTN dynamics.  
If $Z(t) = |\langle \sigma_+ (t) \rangle / \langle \sigma_+ (0) \rangle|$ 
defines the controlled decay function, the following expression holds in 
the presence of a single BC after a time $t =2N\Delta t$ corresponding to 
$N$ control cycles \cite{FaoroNext}:
\begin{eqnarray}
Z(t) &\hspace*{-1.3mm} = \hspace*{-1.3mm}
& a^N \bigg\{  \hspace*{-.5mm} \frac{f(\alpha)+f(-\alpha)}{4} 
\mbox{}_2 F_{1} \hspace*{-1mm}\left [
\frac{1-N}{2},1-\frac{N}{2};1-N;z^2 \right] \nonumber \\
& \hspace*{-1.3mm}- \hspace*{-1.3mm}& |\alpha|^4 \,
\mbox{}_2 F_{1} \hspace*{-1mm}\left [
1-\frac{N}{2},\frac{3}{2}-\frac{N}{2};2-N;z^2\right ] \bigg\} \:.
\hspace*{-3mm}\label{hyper}
\end{eqnarray}
Here, $\mbox{}_2 F_{1}$ denotes the regularized hypergeometric 
function \cite{Stegun}, $\alpha =\sqrt{ 1-g^2 -2i g
\tanh(\epsilon /2 k_B T) }$ = $\alpha' +i \alpha''$, 
$z ={2 e^{- \gamma \Delta t}}/a$, and $a$, $f(\alpha)$ are 
respectively given by:
\begin{eqnarray}
a  = \frac{e^{-\gamma \Delta t}}{|\alpha|^2} \Big[ 
&\hspace*{-2mm} (\hspace*{-2mm} & 1+g^2+|\alpha|^2) 
\cosh(\gamma \alpha' \Delta t) \nonumber \\
-  &\hspace*{-2mm} ( \hspace*{-2mm} & 1+g^2-|\alpha|^2) 
\cos(\gamma \alpha'' \Delta t) \Big] \:, \nonumber 
\end{eqnarray}
\vspace*{-5mm}
\begin{eqnarray}
f(\alpha)&\hspace*{-1mm}  = \hspace*{-1mm} &
e^{\alpha' \Delta t} \Big [|\alpha|^2 +
(1+g^2-2 i g (\delta p_{eq}-\delta p_0)) \nonumber \\
&\hspace*{-1mm}  + \hspace*{-1mm} & 2 
(\alpha'+\delta p_0 g \alpha'') \Big]  
+  e^{i \alpha'' \Delta t} \Big [
2 i(\alpha''+ \delta p_0 g \alpha' ) \nonumber \\
& \hspace*{-1mm} + \hspace*{-1mm} & \hspace*{10mm} |\alpha|^2 -
(1+g^2 -2 i g (\delta p_{eq} -\delta p_0 )) \Big]\:, \nonumber
\end{eqnarray}
$\delta p_{eq} = \tanh(\epsilon /2 k_B T)$ and $\delta p_0= \pm1$ 
denoting the equilibrium and initial value of the BC population 
difference.  
 
For $M$ noise sources, the individual contributions factorize, 
$Z(t)=\prod_k Z_k(t)$.  In the absence of control, the result of 
\cite{Paladino2002} (Eq. (4)) is recovered.  In the continuous limit 
where $\Delta t = t/ N \rightarrow 0$ \cite{Viola1}, one can prove 
using Eq. (2) that $Z_k (t) \rightarrow 1$ $\forall k$, confirming 
perfect decoherence suppression for ideal decoupling.  Thus,
{\em complete} 1/$f$ compensation requires access to control time 
scales $\Delta t \lesssim 1/\gamma_{max}$, so as to quench the influence 
of the {\em fastest} fluctuator present in the environment.  
We find that an almost full coherence recovery (better than 90\%) can always 
be achieved if decoupling is fast relative to the {\em effective upper cutoff}, 
$\Delta t \lesssim 1/\gamma^e_{max}$, implying controls up to an order 
of magnitude slower.  Tighter estimates of the minimum control rates able 
to ensure a significant decoherence suppression are possible upon 
identifying the dominant noise sources in the relevant dynamical regime.  
While a detailed analysis is deferred to \cite{FaoroNext}, 
the salient features may be summarized as follows.

{\it Decoupling in the pure dephasing regime.$-$} Adiabatic decoherence 
is insensitive to the energy scale $\Delta E = \Omega$, but critically 
dependent on the coupling strength distribution, the overall time scale 
of the process being largely determined by $\langle v \rangle$ for small 
dispersions as assumed.  An important consequence of $\Delta E$ being 
irrelevant is that {\em decoherence acceleration} \cite{Viola1} is 
{\em never} observed for $\Delta =0$ with arbitrary control parameters.  
Symmetric decoupling performs systematically better over its 
asymmetric counterpart, differences being larger for a number of 
control cycles $N\lesssim 5$.  For charge-echo sequences with realistic 
non-Gaussian noise and $\Delta t$ values \cite{Nakamura2002}, the 
symmetric readout signal can be up to 70\% higher at times where the free 
coherence has entirely decayed (Fig.~1).  Effects related to temperature 
as well as different initializations of the BCs (including sample to sample 
variations and correlations when repetitions of the readout process are 
needed as in \cite{Nakamura2002}) can be fully taken into account 
\cite{FaoroNext}.  While this may be critical for a quantitative 
comparison with experiments, we anticipate no substantial changes as 
far as the decoupling efficacy is concerned.

\begin{figure}[h]
\psfrag{x}{$\hspace*{-7mm} $ time (ns)}
\psfrag{y}{$\hspace*{-5mm}|\langle \sigma_+ (t) \rangle|$}
\includegraphics[width=2.9in,height=2.7in]{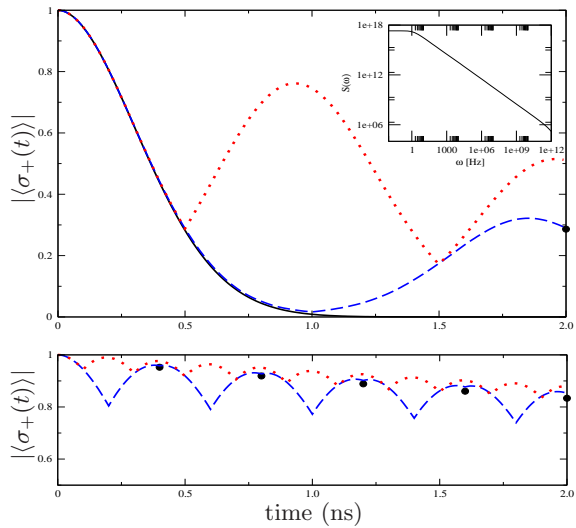}
\caption{ Symmetric (${\cal P}_S$, dotted) vs asymmetric 
(${\cal P}_A$, dashed) decoupling 
for pure dephasing from a realistic 1/$f$ spectrum \cite{Nakamura2002} 
(Inset: $\gamma_{min}$=1 Hz (extrapolated), $\gamma_{max}=10^{12}$ Hz, 
$n_d=1000$, $\langle v\rangle = 9.2 \cdot 10^7$ Hz, 
$\Delta v /\langle v \rangle \sim 0.2$). 
Upper panel: free evolution (solid line) vs echo signals, $\Delta t = 1$ 
ns.  Lower panel: $N=5$, $\Delta t = 0.2$ ns.  Stroboscopic data points 
from Eq. (\ref{hyper}) are shown at $t=2N\Delta t$, whereas continuous 
curves result from averaging over $10^5$ RTN realizations. Each BC is 
initially assumed in a thermal mixture with 
$\epsilon/2 k_B T \simeq \delta p_{eq}= 0.08$.  }
\end{figure}

A key factor determining the control effectiveness 
for dephasing processes is {\em where} $\langle v \rangle$ is 
positioned within $[\gamma_{min}, \gamma_{max}]$. 
If noise is purely Gaussian, $v_k/\gamma_k \ll 1$ $\forall k$, 
then $\langle v \rangle \lesssim \gamma_{min}$ for realistic spectra.  
Gaussian dephasing is dominated by the lowest spectral components, 
as witnessed by the diverging rate predicted by second-order 
perturbation theory for $\omega \rightarrow 0$.  Hence, averaging 
of the slowest decades near $\gamma_{min}$ {\em suffices} 
for a dramatic coherence improvement (Fig.~2(a)).  Non-Gaussian 
behavior may arise either for a lower $\gamma_{min}$ at fixed 
$\langle v \rangle$ (Fig.~2(b)), or for a higher $\langle v \rangle$ 
in a fixed range $[\gamma_{min},\gamma_{max}]$ (Fig.~2(c)).  
Irrespective of whether saturation effects occur in the dynamics of 
non-Gaussian charges \cite{Paladino2002,FaoroNext}, it turns out that 
a substantial decoherence contribution originates from noise decades 
near $\langle v \rangle$, thus making their compensation essential.
In practice, a necessary and sufficient criterion to obtain between 
60-75\% coherence recovery is to choose $\Delta t \lesssim \text{min} 
(1/10^q \gamma_{min}, 1/10^q \langle v \rangle )$, with $q=1$ or $2$ 
at most.  For fixed $\gamma_{min}$, this allows for pulse repetition 
times which can be longer and longer the lower the value of 
$\langle v \rangle$.  In a strongly non-Gaussian regime where the 
latter moves to higher frequencies, the advantages of slow 
decoupling are lost, and control rates ruled by $\gamma^e_{max}, 
\gamma_{max}$ may become necessary.  

\begin{figure}[t]
\psfrag{x}{$\hspace*{-8mm} $ time $(1/\Delta E)$}
\psfrag{y}{$\hspace*{-6mm} |\langle \sigma_+ (t)\rangle |$}
\includegraphics[width=2.8in,height=3.3in]{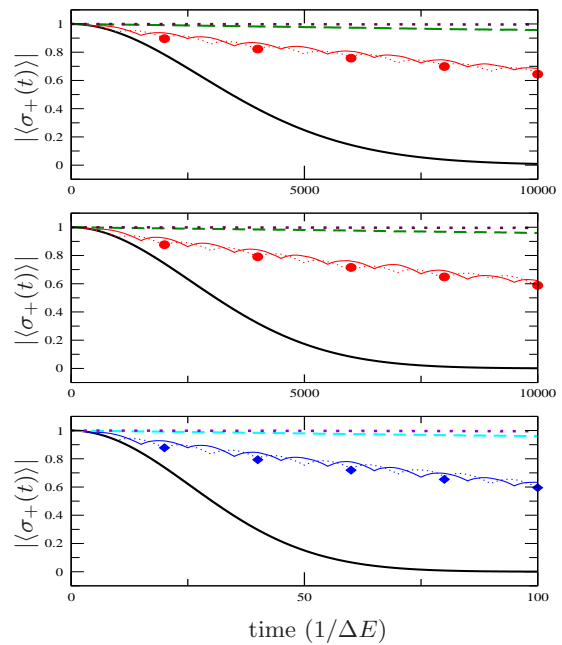}
\caption{1/$f$ suppression under pure dephasing conditions,
$\Omega=1, \Delta=0$.  Top to bottom: 
(a) Purely Gaussian spectrum, $\gamma_{min}=10^{-4}$, 
$\gamma_{max}=100$, $n_d=100$, $\langle v\rangle = 10^{-4}$; 
(b) and (c) Non-Gaussian spectra with parameters as in (a)
except that $\gamma_{min}= 10^{-6}$ in (b), and 
$\langle v \rangle = 0.01$ in (c), respectively. 
$\gamma^e_{max}=10$ in all cases.   
Solid lines depict free evolutions. Control parameters are: 
(a) and (b) $\Delta t = 1000$ (thin solid), 
$\Delta t = 100$ (dashed), $\Delta t = 10$ (dotted);  
(c) $\Delta t = 10$ (thin solid), $\Delta t = 1$ (dashed), 
$\Delta t = 0.1 \sim 1/\gamma^e_{max}$ (dotted).  
Results are averages over $2\cdot 10^4$ RTN realizations
under ${\cal P}_S$ protocols.  Each BC is initially in a pure 
state randomly sampled according to 
$\langle \delta p_0 \rangle= \delta p_{eq}=0.08$. 
Asymmetric RTN and analytic results from Eq. (\ref{hyper}) are 
also shown for $N=5$ cycles. 
}
\end{figure}

{\it Decoupling in the charge degeneracy regime.$-$}  A distinctive 
feature of the dissipative dynamics in this limit is the sensitivity to 
the energy scale $\Delta E=\Delta$.  For Gaussian noise, this may be 
understood from the fact that both relaxation and dephasing are governed 
(up to a factor 2) by a single rate, which depends only on the power 
spectrum $S(\omega=\Delta E)$ to second order in the couplings.  We find 
that the dynamical role of the energy scale $\Delta E$ extends beyond the 
Gaussian limit.  In particular, irrespective of the Gaussian or non-Gaussian 
nature of the spectrum, decoherence rates at charge degeneracy do not 
substantially differ from adiabatic ones if the qubit operates at 
sufficiently low frequencies, $\Delta E \rightarrow \gamma_{min}$.  
Shifting the working point to higher frequency has the effect of 
filtering out the majority of noise contributions, leading to coherence 
times which are orders of magnitude longer than in the corresponding 
dephasing configuration (Fig.~3a).  However, a hallmark 
of control in this regime is the possibility to {\em accelerate} 
decoherence if cycle times long with respect to the free period 
$2\pi/\Delta E$ are used.  This is evident in the oscillatory 
dynamics of $\langle \sigma_z (t) \rangle$ (Figs.~3b--d). If 
$\Delta t \gg \pi/\Delta E$, the improvement in the controlled 
amplitude does not compensate for the effective decay due to the 
frequency mismatch relative to the free oscillations, and overall 
acceleration results \cite{FaoroNext}.  As $\Delta t$ approaches 
$\pi/\Delta E$ from above, the relative importance of these two 
mechanisms reverses, and a crossover to noise suppression occurs at 
times which are shorter the shorter $\Delta t$. Full coherence 
recovery is guaranteed for $\Delta t \lesssim 1/10 \Delta E$ which, 
however, may be close to $1/\gamma^e_{max}$ if $\Delta E$ is large. 
  
\begin{figure}[t]
\psfrag{x}{$\hspace*{-8mm} $ time $(1/\Delta E)$}
\psfrag{y}{$\hspace*{-4mm}\langle \sigma_z (t) \rangle$}
\psfrag{z}{$\hspace*{-4mm}|\langle \sigma_+ (t) \rangle|$}
\includegraphics[width=2.9in,height=3.4in]{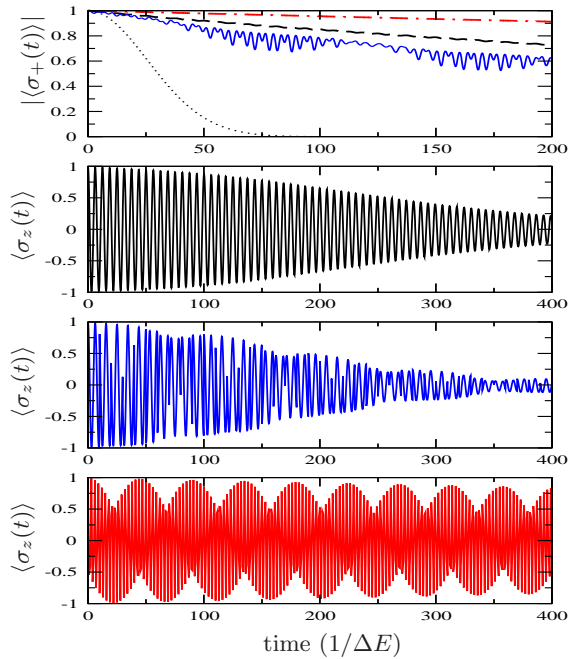}
\caption{1/$f$ suppression under charge degeneracy conditions, 
$\Omega=0, \Delta=1$.  Noise spectrum as in Fig.~2(c).  
Top panel, (a): coherence in the $z$ basis for free (dashed) 
and controlled dynamics, $\Delta t = 10$ (solid), $\Delta t= 1$ 
(dash dotted). The dephasing curve at $\Delta E=1$ is reproduced 
from Fig. 2(c) (dotted).  Note the deteriorated performance for 
$\Delta t = 10$.  Lower panels: damping (b, no control), acceleration 
(c, $\Delta t=10$), and partial recovery (d, $\Delta t =1$) of
coherent charge oscillations.  Complete recovery (not shown) is 
found for $\Delta t=0.1 \sim 1/\gamma^e_{max}$.
RTN averages (${\cal P}_S$ protocols) are taken over $2\cdot 10^5$ 
realizations, with environment initialized as in Fig.~2.  
}
\end{figure}

Interestingly, a trade-off emerges in the charge degeneracy regime between 
noise effects which are stronger but easier to decouple (as $\Delta E$ 
shifts toward low frequencies); and noise effects which are substantially 
reduced from the beginning but harder to suppress (as $\Delta E$ increases).  
A similar conclusion is likely to hold for generic qubit parameters. Thus, 
while it may seem counterproductive to consider low operation frequencies 
in the presence of 1/$f$ noise, in practice both the relevant noise level 
and the available control resources should guide the choice of a working 
point able to maximize the control efficiency. 
 
{\it Conclusion.$-$} We have characterized the performance of decoupling 
techniques at reducing 1/$f$ noise based on a realistic model.  Beside 
identifying noise scenarios where decoupling is highly effective with 
affordable rates, our results suggest the possibility of using 
control as a {\em diagnostic tool} to infer spectral properties (such as 
$\langle v\rangle$ or $\gamma^e_{max}$), which are not directly measurable.  
While actual details will be device-dependent and require a dedicated 
analysis incorporating the underlying physics, we expect that our main
conclusions will have a wide range of applicability, including noise 
spectra with power-law behavior of the form 1/$|\omega|^\alpha$ 
with $\alpha >1$ \cite{Weissman1988}, or alternative qubit design as 
demonstrated in \cite{Vion}.  Ultimately, our results hold the promise 
that decoupling methods may substantially improve the prospects for 
reliable solid-state QIP.

\acknowledgments 
L.F. acknowledges support from Los Alamos National Laboratory, where 
this work was initiated, and from the EU(IST-FET-SQUBIT). It is a pleasure 
to thank P. Giorda, E. Knill, and R. Onofrio for discussions.

\vspace*{-2mm}

\bibliography{/n/u2/lviola/papers/qip}

\begin{thebibliography}{21}
\expandafter\ifx\csname natexlab\endcsname\relax\def\natexlab#1{#1}\fi
\expandafter\ifx\csname bibnamefont\endcsname\relax
  \def\bibnamefont#1{#1}\fi
\expandafter\ifx\csname bibfnamefont\endcsname\relax
  \def\bibfnamefont#1{#1}\fi
\expandafter\ifx\csname citenamefont\endcsname\relax
  \def\citenamefont#1{#1}\fi
\expandafter\ifx\csname url\endcsname\relax
  \def\url#1{\texttt{#1}}\fi
\expandafter\ifx\csname urlprefix\endcsname\relax\def\urlprefix{URL }\fi
\providecommand{\bibinfo}[2]{#2}
\providecommand{\eprint}[2][]{\url{#2}}

\bibitem[{\citenamefont{Weissman}(1988)}]{Weissman1988}
\bibinfo{author}{\bibfnamefont{M.~B.} \bibnamefont{Weissman}},
  \bibinfo{journal}{Rev. Mod. Phys.} \textbf{\bibinfo{volume}{60}},
  \bibinfo{pages}{537} (\bibinfo{year}{1988}).

\bibitem[{\citenamefont{Press}(1978)}]{Press1978}
\bibinfo{author}{\bibfnamefont{W.~H.} \bibnamefont{Press}},
  \bibinfo{journal}{Astrophys.} \textbf{\bibinfo{volume}{7}},
  \bibinfo{pages}{103} (\bibinfo{year}{1978}).

\bibitem[{\citenamefont{Voss}(1992)}]{Voss1992}
\bibinfo{author}{\bibfnamefont{R.~F.} \bibnamefont{Voss}},
  \bibinfo{journal}{Phys. Rev. Lett.} \textbf{\bibinfo{volume}{68}},
  \bibinfo{pages}{3805} (\bibinfo{year}{1992}).

\bibitem[{\citenamefont{Lo}(1991)}]{Lo1991}
\bibinfo{author}{\bibfnamefont{A.~W.} \bibnamefont{Lo}},
  \bibinfo{journal}{Econometrica} \textbf{\bibinfo{volume}{59}},
  \bibinfo{pages}{1279} (\bibinfo{year}{1991}).

\bibitem[{Spe()}]{SpecialIssue}
\bibinfo{note}{Special Focus Issue on {\it Experimental Proposals for Quantum
  Computation}, Fortschr. Phys. {\bf 48} (2000)}.

\bibitem[{\citenamefont{Covington et~al.}(2000)\citenamefont{Covington, Keller,
  Kautza, and Martinis}}]{Covington2000}
\bibinfo{author}{\bibfnamefont{M.}~\bibnamefont{Covington}},
  \bibinfo{author}{\bibfnamefont{M.~W.} \bibnamefont{Keller}},
  \bibinfo{author}{\bibfnamefont{R.~L.} \bibnamefont{Kautza}},
  \bibnamefont{and} \bibinfo{author}{\bibfnamefont{J.}~\bibnamefont{Martinis}},
  \bibinfo{journal}{Phys. Rev. Lett.} \textbf{\bibinfo{volume}{84}},
  \bibinfo{pages}{5192} (\bibinfo{year}{2000}).

\bibitem[{\citenamefont{Nakamura et~al.}(1999)\citenamefont{Nakamura, Pashkin,
  and Tsai}}]{Nakamura1999}
\bibinfo{author}{\bibfnamefont{Y.}~\bibnamefont{Nakamura}},
  \bibinfo{author}{\bibfnamefont{Y.~A.} \bibnamefont{Pashkin}},
  \bibnamefont{and} \bibinfo{author}{\bibfnamefont{J.~S.} \bibnamefont{Tsai}},
  \bibinfo{journal}{Nature (London)} \textbf{\bibinfo{volume}{398}},
  \bibinfo{pages}{786} (\bibinfo{year}{1999}).

\bibitem[{\citenamefont{Nakamura et~al.}(2002)\citenamefont{Nakamura, Pashkin,
  Yamamoto, and Tsai}}]{Nakamura2002}
\bibinfo{author}{\bibfnamefont{Y.}~\bibnamefont{Nakamura}},
  \bibinfo{author}{\bibfnamefont{Y.~A.} \bibnamefont{Pashkin}},
  \bibinfo{author}{\bibfnamefont{T.}~\bibnamefont{Yamamoto}}, \bibnamefont{and}
  \bibinfo{author}{\bibfnamefont{J.~S.} \bibnamefont{Tsai}},
  \bibinfo{journal}{Phys. Rev. Lett.} \textbf{\bibinfo{volume}{88}},
  \bibinfo{pages}{047901} (\bibinfo{year}{2002}).

\bibitem[{Mar()}]{Martinis}
\bibinfo{note}{J. M. Martinis {\it et al.}, Phys. Rev. B {\bf 67}, 094510
  (2003).}

\bibitem[{\citenamefont{Shiokawa and Lidar}(2002)}]{Shiokawa2002}
\bibinfo{author}{\bibfnamefont{K.}~\bibnamefont{Shiokawa}} \bibnamefont{and}
  \bibinfo{author}{\bibfnamefont{D.~A.} \bibnamefont{Lidar}},
  \bibinfo{journal}{quant-ph/0211081}  (\bibinfo{year}{2002}).

\bibitem[{\citenamefont{Shnirman et~al.}(2002)\citenamefont{Shnirman, Makhlin,
  and Sch\"on}}]{Shnirman2002}
\bibinfo{author}{\bibfnamefont{A.}~\bibnamefont{Shnirman}},
  \bibinfo{author}{\bibfnamefont{Y.}~\bibnamefont{Makhlin}}, \bibnamefont{and}
  \bibinfo{author}{\bibfnamefont{G.}~\bibnamefont{Sch\"on}},
  \bibinfo{journal}{Physica Scripta} \textbf{\bibinfo{volume}{T102}},
  \bibinfo{pages}{147} (\bibinfo{year}{2002}).

\bibitem[{\citenamefont{Paladino et~al.}(2002)\citenamefont{Paladino, Faoro,
  Falci, and Fazio}}]{Paladino2002}
\bibinfo{author}{\bibfnamefont{E.}~\bibnamefont{Paladino}},
  \bibinfo{author}{\bibfnamefont{L.}~\bibnamefont{Faoro}},
  \bibinfo{author}{\bibfnamefont{G.}~\bibnamefont{Falci}}, \bibnamefont{and}
  \bibinfo{author}{\bibfnamefont{R.}~\bibnamefont{Fazio}},
  \bibinfo{journal}{Phys. Rev. Lett.} \textbf{\bibinfo{volume}{88}},
  \bibinfo{pages}{228304} (\bibinfo{year}{2002}).

\bibitem[{\citenamefont{Gutmann et~al.}(2003)\citenamefont{Gutmann, Wilhelm,
  Kaminsky, and Lloyd}}]{Gutmann2003}
\bibinfo{author}{\bibfnamefont{H.}~\bibnamefont{Gutmann}},
  \bibinfo{author}{\bibfnamefont{F.~K.} \bibnamefont{Wilhelm}},
  \bibinfo{author}{\bibfnamefont{W.~M.} \bibnamefont{Kaminsky}},
  \bibnamefont{and} \bibinfo{author}{\bibfnamefont{S.}~\bibnamefont{Lloyd}},
  \bibinfo{journal}{quant-ph/0308107}  (\bibinfo{year}{2003}).

\bibitem[{\citenamefont{Dutta and Horn}(1981)}]{Dutta1981}
\bibinfo{author}{\bibfnamefont{P.}~\bibnamefont{Dutta}} \bibnamefont{and}
  \bibinfo{author}{\bibfnamefont{P.~M.} \bibnamefont{Horn}},
  \bibinfo{journal}{Rev. Mod. Phys.} \textbf{\bibinfo{volume}{53}},
  \bibinfo{pages}{497} (\bibinfo{year}{1981}).

\bibitem[{\citenamefont{Haeberlen}(1976)}]{HaeberlenBook}
\bibinfo{author}{\bibfnamefont{U.}~\bibnamefont{Haeberlen}},
  \emph{\bibinfo{title}{High Resolution NMR in Solids: Selective Averaging}}
  (\bibinfo{publisher}{Academic Press}, \bibinfo{address}{New York},
  \bibinfo{year}{1976}).

\bibitem[{Vio({\natexlab{a}})}]{Viola1}
\bibinfo{note}{L. Viola and S. Lloyd, Phys. Rev. A {\bf 58}, 2733 (1998); L.
  Viola, E. Knill, and S. Lloyd, Phys. Rev. Lett. {\bf 82}, 2417 (1999); L.
  Viola and E. Knill, Phys. Rev. Lett. {\bf 90}, 037901 (2003).}

\bibitem[{\citenamefont{Bauernschmitt and Nazarov}(1993)}]{Bauernschmitt1993}
\bibinfo{author}{\bibfnamefont{R.}~\bibnamefont{Bauernschmitt}}
  \bibnamefont{and} \bibinfo{author}{\bibfnamefont{Y.~V.}
  \bibnamefont{Nazarov}}, \bibinfo{journal}{Phys. Rev. B}
  \textbf{\bibinfo{volume}{47}}, \bibinfo{pages}{9997} (\bibinfo{year}{1993}).

\bibitem[{\citenamefont{Gheorghiu-Svirschevski}(2002)}]{Gheorghiu}
\bibinfo{author}{\bibfnamefont{S.}~\bibnamefont{Gheorghiu-Svirschevski}},
  \bibinfo{journal}{Phys. Rev. A} \textbf{\bibinfo{volume}{66}},
  \bibinfo{pages}{032101} (\bibinfo{year}{2002}).

\bibitem[{\citenamefont{Faoro and Viola}()}]{FaoroNext}
\bibinfo{author}{\bibfnamefont{L.}~\bibnamefont{Faoro}} \bibnamefont{and}
  \bibinfo{author}{\bibfnamefont{L.}~\bibnamefont{Viola}}, \bibinfo{note}{in
  preparation}.

\bibitem[{\citenamefont{Abramowitz and Stegun}(1970)}]{Stegun}
\bibinfo{author}{\bibfnamefont{M.}~\bibnamefont{Abramowitz}} \bibnamefont{and}
  \bibinfo{author}{\bibfnamefont{I.~A.} \bibnamefont{Stegun}},
  \emph{\bibinfo{title}{Handbook of Mathematical Functions}}
  (\bibinfo{publisher}{Dover}, \bibinfo{address}{{New York}},
  \bibinfo{year}{1970}), \bibinfo{note}{p. 556}.

\bibitem[{Vio({\natexlab{b}})}]{Vion}
\bibinfo{note}{D. Vion {\it et al.}, Science {\bf 296}, 886 (2002).}

\end{thebibliography}

\end{document}